\documentclass[review]{elsarticle}

\usepackage[utf8]{inputenc}
\usepackage{lineno, hyperref}
\modulolinenumbers[1]
\usepackage{amsmath}
\usepackage{amssymb}
\usepackage{booktabs}

\makeatletter
\def\ps@pprintTitle{%
  \let\@oddhead\@empty
  \let\@evenhead\@empty
  \def\@oddfoot{\hfill\thepage\hfill}%
  \def\@evenfoot{\hfill\thepage\hfill}%
}
\makeatother

\journal{Journal of \LaTeX\ Templates}

\begin{document}
\pagestyle{plain}

\begin{frontmatter}

\title{Compressing Large Language Models with PCA Without Performance Loss}
\tnotetext[mytitlenote]{Based on the elsarticle package available on \href{http://www.ctan.org/tex-archive/macros/latex/contrib/elsarticle}{CTAN}.}

\author[1]{Magnus Bengtsson\fnref{myfootnote1}}
\address[1]{Department of Engineering, University of Borås, Sweden}
\fntext[myfootnote1]{Corresponding author: magnus.bengtsson@hb.se}
\begin{abstract}
We demonstrate that Principal Component Analysis (PCA), when applied in a structured manner—either to polar-transformed images or segment-wise to token sequences—enables extreme compression of neural models without sacrificing performance. Across three case studies, we show that:

\begin{itemize}
    \item A one-layer classifier trained on PCA-compressed polar MNIST achieves over 98\% accuracy using only 840 parameters.
    \item A 2-layer transformer trained on 70-dimensional PCA-reduced MiniLM embeddings reaches 76.62\% accuracy on the 20 Newsgroups dataset with just 81k parameters.
    \item A decoder-only transformer generates coherent token sequences from 70-dimensional PCA embeddings while preserving over 97\% cosine similarity with full MiniLM representations, using less than 17\% of GPT-2’s parameter count.
\end{itemize}

These results highlight PCA-based input compression as a general and effective strategy for aligning model capacity with information content, enabling lightweight architectures across multiple modalities.
\end{abstract}

\begin{keyword}
\texttt{elsarticle.cls}\sep \LaTeX\sep Elsevier \sep template, preprint 
\MSC[2010] 00-01\sep 99-00
\end{keyword}
\end{frontmatter}
\section{Introduction}

Large neural models offer high performance but often suffer from inefficiency and redundancy. Their input spaces are high-dimensional, their architectures overparameterized, and their capacity often wasted on encoding noise or variation that is irrelevant to the task. This paper takes a different stance—starting not from model scale, but from data structure.

We propose that the apparent need for scale arises from a mismatch between raw representations and the intrinsic information content required to solve the task. Images, text, and sequences all exhibit internal structure—topological, geometric, or sequential—which is frequently ignored in favor of unfiltered embeddings \cite{bengtsson2025c2gkd}. This leads to models that learn more by seeing too much, not by understanding better.

Our method centers on a simple, universal tool: Principal Component Analysis (PCA) \cite{bengtsson2025pca,pearson1901pca}. Applied globally (e.g., to polar-transformed image coordinates) or locally (e.g., per-token in a text sequence), PCA compresses raw input into a lower-dimensional latent space that preserves semantic structure. These PCA components function as information-rich bottlenecks that filter away variation irrelevant to the task.

\paragraph{Topology First, Typology Later.\cite{bengtsson2025c2gkd}}
In most machine learning systems, data is treated as a syntactic object—arrays of numbers destined for classification. But natural data manifests first as structure: shapes, patterns, sequences. These are \textit{semantic in themselves}—recognizable before they are labeled. Labels—what we call typology—are imposed after the fact, as a symbolic overlay on a fundamentally topological world.

We adopt a \textbf{structure-first} perspective. PCA serves not just as a dimensionality reducer, but as a semantic encoder—extracting class-defining structure from the input before labels are even considered. Models trained on these representations learn to operate in a space where the signal has already been distilled.

Principal Component Analysis (PCA) has long been recognized as an effective technique for dimensionality reduction and pattern recognition. One of the earliest and most influential applications is the use of \emph{eigenfaces} for face recognition \cite{turk1991eigenfaces}, where PCA was applied to a dataset of facial images to extract the most salient variation directions—capturing identity-relevant features in a compact linear subspace.

The success of this approach in visual classification tasks demonstrated that carefully structured PCA representations can retain class-discriminative information even under aggressive compression. This insight has inspired further work in our prior research \cite{bengtsson2025c2gkd}, where PCA was used as a generative constraint to synthesize structurally valid synthetic data in the absence of real samples.

Transformer-based models \cite{vaswani2023attentionneed} have become the dominant architecture in natural language processing due to their ability to model long-range dependencies and scale efficiently. A key precursor to these models is the use of distributed word representations, such as those introduced in the \emph{word2vec} framework \cite{mikolov2013efficientestimationwordrepresentations}, which enabled the mapping of discrete tokens to dense, semantically meaningful vector spaces. This embedding step remains a foundational component in modern transformer pipelines, where each token is first converted into a fixed-length vector prior to further processing.

\paragraph{Compression Before Training}
Rather than reducing model size post hoc, we compress the input space \textit{before} training. This allows us to train drastically smaller models from the outset, without sacrificing performance. We demonstrate this strategy in three domains:
\begin{itemize}
\item \textbf{Image classification:} PCA on polar-transformed MNIST \cite{lecun1998gradient} images enables a one layer fully connected network \cite{bishop1995neural,rumelhart1986learning} variant with only 840 parameters to reach 98\% accuracy.
    \item \textbf{Text classification:} PCA applied to token-wise MiniLM embeddings reduces dimensionality from 384 to 70, with a two-layer transformer retaining 76.62\% accuracy on 20 Newsgroups.
    \item \textbf{Text generation:} A GPT-style decoder trained on PCA-compressed embeddings preserves semantic coherence with less than 17\% of the parameter count of standard models.
\end{itemize}

\paragraph{An Information-Theoretic View}
Our use of PCA aligns naturally with Shannon’s view of communication \cite{shannon1948mathematical}: a model can be treated as a channel with limited capacity, and the data as a compressed message. The number of model parameters defines the channel’s bandwidth.

Let $k$ be the number of PCA components needed to faithfully represent the data distribution. Then the model must have sufficient capacity to encode, manipulate, and decode variation in this $k$-dimensional latent space. This leads to a central insight: \textit{model size should scale with the intrinsic complexity of the task—not with the raw dimensionality of the input}.

In this view, PCA serves a role analogous to RNA in biological systems: a structured, repeatable, information-preserving intermediate form. Training becomes less about brute-force memorization, and more about shaping a transmission channel that matches the entropy of the compressed data

\paragraph{Purpose and Contribution.}
This article aims to demonstrate that PCA-based input compression—whether applied globally to image coordinates or locally to token embeddings—can dramatically reduce the number of trainable parameters required for effective learning. By restructuring the input space to reflect the intrinsic information content of the data, we show that smaller models can perform on par with, or even surpass, significantly larger counterparts.

The results provide strong empirical evidence that principled compression is not a compromise, but a path to efficiency. When the input representation is aligned with the underlying task structure, fewer parameters are needed—not because the model is weaker, but because the data has already done part of the work.

\section*{A Hierarchical View of Token Compression}

\subsection*{Token Vectors as Carriers of Semantics}

At the heart of every transformer lies a deceptively simple construct: the token embedding vector. These high-dimensional vectors, often initialized through models such as word2vec \cite{mikolov2013efficientestimationwordrepresentations} or BERT \cite{devlin2019}, form the foundation of symbolic representation in neural language models. Each vector encapsulates not a single meaning, but a spectrum of semantic affordances—syntactic roles, contextual tendencies, grammatical associations—all compressed into a single point in $\mathbb{R}^D$.

The philosophy behind these vectors is distributional: ``You shall know a word by the company it keeps.'' \cite{firth1957synopsis}. But unlike traditional feature engineering, these vectors are not handcrafted—they are learned through co-occurrence, prediction, and contrastive structure. Their dimensionality, typically 300--1024, is not arbitrary, but chosen to preserve separability and relational structure.

\subsection{Why High-Dimensional Vectors Work}

From a geometric standpoint, a high-dimensional vector space allows each token to be represented as a nearly orthogonal point in a sparse semantic lattice. The curse of dimensionality becomes, paradoxically, a blessing: with enough dimensions, vectors can be placed far enough apart to encode both local and global relationships.

Yet, this abundance comes at a cost: model capacity, memory bandwidth, and trainability. Not all dimensions carry equal semantic weight, and some may encode redundant or task-irrelevant variation. Recent works have shown that a surprisingly small number of principal axes often account for most of the semantic variance \cite{gong2018frage,may2019downstream,huertas2022dimred}.

This observation prompts a deeper question: what is the minimal dimensionality required to preserve a token's identity across downstream tasks? Put differently—how many dimensions are needed to distinguish a token not in isolation, but in its full contextual role?

\subsection{Dimensionality and Discriminability: Theoretical Bounds}

Several approaches have attempted to quantify the required dimensionality for token discriminability. From information-theoretic perspectives \cite{shannon1948mathematical}, to intrinsic dimensionality estimation \cite{levina2005maximum}, and empirical studies of embedding compression \cite{zhang2022kbcompress}, the consensus is emerging: embeddings are highly compressible—yet only up to a point.

The intrinsic dimension of token spaces has been estimated to lie in the range 20--100 depending on corpus, language, and task. This suggests that embeddings of size 384 or 768, as found in MiniLM and BERT, are overcomplete bases—dense, but redundant.

\subsection{PCA: From Measurement to Compression}

Principal Component Analysis (PCA), originally introduced by Pearson \cite{pearson1901pca} and later formalized by Hotelling \cite{hotelling1933analysis}, offers a natural method for identifying and extracting dominant axes of variation. In the context of token embeddings, PCA can be seen as a semantic distillation mechanism: removing dimensions that carry less inter-token discriminative power.

While PCA has been used extensively to visualize \cite{mikolov2013efficientestimationwordrepresentations}, cluster \cite{jolliffe2016pca}, and analyze embedding spaces \cite{may2019downstream,wang2019single}, its use as a **preprocessing step before training a transformer model** remains rare. Our work takes this next step: not just studying PCA, but building around it.

\subsection{A Structural Shift: Compression Before Computation}

Rather than treat compression as an afterthought—applied post-training or as a downstream optimization—we shift the compression upstream. We begin not with overcomplete token vectors, but with their compressed counterparts, projected through PCA bases trained per token position or per document.

This flips the conventional paradigm: the burden of representation lies not in the model, but in the preprocessing. The model no longer learns to ignore redundant features—it never sees them. This architectural stance aligns with prior work in *structure-first* modeling \cite{bengtsson2025c2gkd}, where topology precedes typology, and input alignment reduces the need for brute-force scale.

The remainder of this paper formalizes this perspective, evaluates its efficacy across modalities, and positions it within a broader theoretical framework rooted in information theory, biological replication, and structural representation learning.

\section{Token-to-Embedding Compression Pipeline}

To understand how PCA enables model compression in language tasks, we first describe the full pipeline from raw text to compressed embeddings. This step is essential to contextualize how a 100-token text segment is transformed into a numerical representation used for training our decoder.

\subsection{Tokenization}

Raw text is first split into subword tokens using a byte pair encoding (BPE) tokenizer, such as the one used by GPT-2. For instance, the Swedish word \textit{"kanske"} might be tokenized into ['kan', 'ske'], each corresponding to an entry in the tokenizer’s vocabulary.

\begin{quote}
\textbf{Example:}  
Input text: \texttt{"It might rain"}  
Tokenized: \texttt{["It", " might", " rain"]}
\end{quote}

Each token is represented internally as an integer index into a vocabulary of size $V = 50257$ (in our case, GPT-2's vocabulary).

\subsection{Embedding Layer}

These token indices are then mapped to continuous high-dimensional vectors using a pretrained embedding model such as MiniLM. This produces a matrix $X \in \mathbb{R}^{T \times D}$, where $T$ is the sequence length (e.g., 100 tokens), and $D$ is the embedding dimension (typically 384 for MiniLM).

\[
X = \texttt{Embed}(\texttt{Tokens}) \in \mathbb{R}^{T \times D}
\]

This matrix contains the semantic representation of the input sequence, where each row corresponds to a token in context.
\subsection{Construction of Token Embeddings via MiniLM}

Each token embedding $\mathbf{x}_i \in \mathbb{R}^D$ is computed using a pretrained transformer model, such as MiniLM. This model is a multi-layer neural network that processes the full token sequence and outputs contextualized vectors for each token.

\paragraph{Step 1: Token Embedding Lookup.}
Initially, each token index is mapped to a static embedding vector via a learnable embedding matrix $E \in \mathbb{R}^{V \times D}$, where $V$ is the vocabulary size and $D$ the embedding dimension.

\[
\mathbf{e}_i = E[\texttt{token}_i]
\]

\paragraph{Step 2: Positional Encoding.}
To incorporate word order, a positional encoding vector $\mathbf{p}_i$ is added to each token embedding:

\[
\mathbf{z}_i = \mathbf{e}_i + \mathbf{p}_i
\]

These $\mathbf{z}_i$ vectors form the input to the transformer layers.

\paragraph{Step 3: Contextual Encoding via Transformer.}
The sequence of position-enhanced embeddings $\{\mathbf{z}_1, \dots, \mathbf{z}_T\}$ is passed through $L$ self-attention layers. Each layer refines the representation by allowing each token to attend to others in the sequence.

Let $f_{\text{MiniLM}}$ represent the full transformer model. The final output for token $i$ is:

\[
\mathbf{x}_i = f_{\text{MiniLM}}(\mathbf{z}_i; \mathbf{z}_{1:T}) \in \mathbb{R}^D
\]

This embedding $\mathbf{x}_i$ captures the token’s semantic meaning in the context of the entire sequence.

\paragraph{Step 4: Sequence Embedding Matrix.}
Stacking all token embeddings yields the final embedding matrix:

\[
X = 
\begin{bmatrix}
\mathbf{x}_1^\top \\
\vdots \\
\mathbf{x}_T^\top
\end{bmatrix}
\in \mathbb{R}^{T \times D}
\]

This matrix is the input to the PCA compression described in the next section.

\subsection{Embedding Table and Token Mapping}

Let $\mathcal{V}$ denote the vocabulary of subword tokens used by the tokenizer, with size $|\mathcal{V}| = V$. Each token $w \in \mathcal{V}$ is assigned a unique integer index $t \in \{0, 1, \dots, V-1\}$. 

We define the embedding table as a matrix $E \in \mathbb{R}^{V \times D}$, where each row $E[t] = \mathbf{e}_t \in \mathbb{R}^D$ represents the static embedding vector associated with token index $t$.

The mapping from token to embedding is thus a deterministic function:
\[
\texttt{embed} : \{0, 1, \dots, V-1\} \rightarrow \mathbb{R}^D, \quad \texttt{embed}(t) = \mathbf{e}_t = E[t]
\]

This function is fixed and consistent for a given pretrained model. That is, for a fixed vocabulary and model state, the mapping $t \mapsto \mathbf{e}_t$ is invariant: the same token index will always yield the same embedding vector prior to any contextual transformation.

These embeddings $\mathbf{e}_t$ form the basis for downstream contextualization via transformer layers.
\subsection{PCA Compression per Token Position}

To reduce the dimensionality of this sequence while preserving positional semantics, we apply PCA separately at each token position across many documents. That is, we collect all embeddings at position $i$ across the dataset and perform PCA to find the most representative $k$-dimensional subspace.

Let $X_i \in \mathbb{R}^{N \times D}$ be the collection of embeddings at position $i$ across $N$ sequences. PCA yields:

\[
X_i^{(PCA)} = (X_i - \mu_i) W_i^{\top}, \quad W_i \in \mathbb{R}^{k \times D}
\]

This process yields $T$ separate PCA transforms, each mapping $D$-dimensional embeddings to a compressed $k$-dimensional space (e.g., $k=70$).

\subsection{Resulting Compressed Input}

After PCA, each sequence is represented as a matrix $Z \in \mathbb{R}^{T \times k}$. This matrix serves as the input to our lightweight transformer decoder described in later sections.

\[
Z = \texttt{PCA}(\texttt{Embed}(\texttt{Tokens})) \in \mathbb{R}^{T \times k}
\]

This pipeline transforms natural language into an information-preserving compressed representation with drastically reduced dimensionality, forming the basis for our decoder models.

\section{Model Architectures}

In this section, we present the three model architectures used in our experiments, each tailored to a different data modality—images, text classification, and sequence generation. All models are trained exclusively on PCA-compressed inputs, as described in the previous section.

\subsection{Case 1: Polar-PCA Classifier on MNIST}

To evaluate the effect of input compression on image classification, we construct a minimal fully connected network trained on PCA-compressed representations of MNIST digits.

Each image is first transformed from Cartesian to polar coordinates. The polar image is divided into 28 angular segments, corresponding to the original image rows. For each segment, we apply PCA across the dataset and retain the first three principal components. This results in a compressed input vector of dimension $28 \times 3 = 84$ per image.

The classifier consists of a single fully connected layer with 84 input units and 10 output units, using softmax activation. The total number of trainable parameters is 840.

Training is performed on 10,000 samples from the MNIST training set using standard cross-entropy loss. All experiments were conducted on an NVIDIA L4 GPU.

\subsection{Case 2: Segment-Wise Transformer for Text Classification}

To evaluate PCA-based compression on textual inputs, we construct a lightweight transformer classifier trained on compressed token embeddings from the 20 Newsgroups dataset.

Each document is divided into 100-token blocks and embedded using a pretrained MiniLM model, producing a sequence of token vectors $X \in \mathbb{R}^{100 \times 384}$. To reduce dimensionality, we apply PCA independently at each token position across the dataset. For each of the 100 positions, we retain the top $k = 70$ principal components, resulting in a compressed input representation $Z \in \mathbb{R}^{100 \times 70}$.

The classification model consists of a 2-layer transformer encoder with the following structure:

\begin{itemize}
    \item Input: $Z \in \mathbb{R}^{100 \times 70}$
    \item Two encoder layers with 2 attention heads per layer
    \item Output: mean-pooled sequence → feedforward layer → softmax over 20 classes
\end{itemize}

The total number of trainable parameters in this model is approximately 81,000. Training was performed using standard cross-entropy loss, with positional encoding added before input to the encoder. All experiments were conducted on an NVIDIA L4 GPU.

\subsection{Case 3: TinyGPTDecoder for Autoregressive Generation}

To evaluate PCA-based input compression in a generative context, we construct a lightweight decoder-only transformer trained on compressed MiniLM embeddings.

Input sequences are segmented into blocks of 100 tokens and embedded using a pretrained MiniLM model. The resulting matrix $X \in \mathbb{R}^{100 \times 384}$ is compressed via PCA applied independently at each token position, retaining 70 components per position. This results in an input matrix $Z \in \mathbb{R}^{100 \times k}$, where $k \in \{70\}$.

The model follows a simplified GPT-style decoder architecture:

\begin{itemize}
    \item Input: $Z$ with fixed sinusoidal positional encoding added.
    \item Two causal transformer decoder layers, each with 2 attention heads.
    \item Output: linear projection to GPT-2 vocabulary size ($V = 50257$).
\end{itemize}

The model is trained to autoregressively predict the next token in a sequence using cross-entropy loss. A causal attention mask ensures left-to-right decoding behavior. All training was conducted on an NVIDIA L4 GPU.

\section{Results}

We report experimental results from all three case studies—image classification, text classification, and autoregressive generation—where PCA-compressed inputs are used to train lightweight models with competitive performance.

\subsection{Case 1: Polar-PCA Classifier on MNIST}

The 1-layer fully connected classifier, trained on 84-dimensional PCA vectors derived from polar-transformed MNIST digits, achieves over 98\% accuracy on the test set. The model contains only 840 trainable parameters.

This demonstrates that even extremely simple architectures can achieve high performance when inputs are semantically structured. Given that 10,000 training samples are used, the result highlights a substantial redundancy in the dataset: most samples reinforce known structures rather than adding new information.

To assess the effect of training data volume on model performance and computational cost, we conducted a controlled experiment varying the number of training samples used to fit a fixed model architecture. Accuracy was measured on a held-out test set, while total training time was logged in seconds, including forward and backward passes for all epochs. The results shown in Figure \ref{fig:accuracy_vs_data} illustrate the diminishing returns in accuracy with increasing dataset size, as well as the expected near-linear growth in training time.

\begin{figure}[h]
    \centering
    \includegraphics[width=0.75\linewidth]{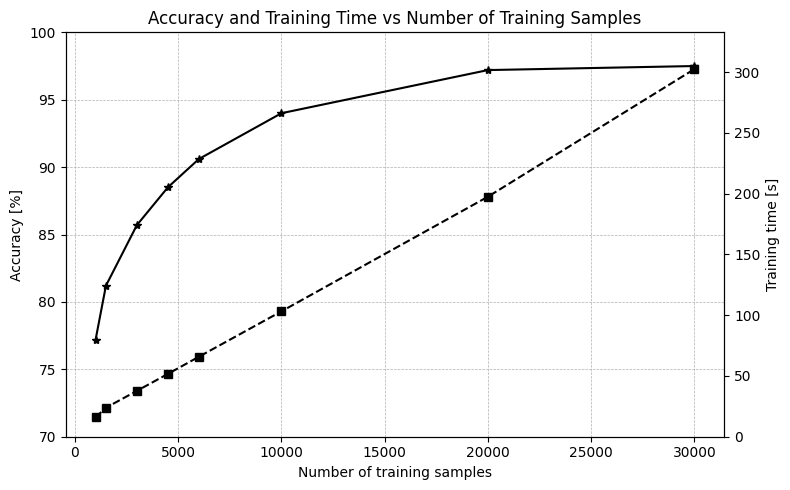}
    \caption{Relationship between the number of training samples, classification accuracy, and training time. The left y-axis shows the model’s accuracy on a held-out test set, while the right y-axis shows the total training time required to reach convergence. Accuracy gains taper off beyond 10,000 samples, indicating saturation, while training time continues to increase almost linearly with dataset size.}
    \label{fig:placeholder}
\end{figure}

\subsection{Case 2: PCA-Constrained Transformer on 20 Newsgroups}

Our compact transformer classifier, trained on PCA-reduced MiniLM embeddings ($100 \times 70$), achieves 76.62\% accuracy on the 20 Newsgroups test set. This is notable given the model's size of only 81,000 parameters—orders of magnitude smaller than BERT-based alternatives.

\begin{table}[h]
\centering
\begin{tabular}{lccc}
\toprule
\textbf{Model} & \textbf{Input Dim.} & \textbf{Parameters} & \textbf{Test Accuracy} \\
\midrule
BERT-base (fine-tuned) & $512 \times 768$ & $\sim$110M & 86--88\% \\
MiniLM + FFN & $100 \times 384$ & $\sim$2M & 76--78\% \\
\textbf{Ours (PCA-Transformer)} & $100 \times 70$ & \textbf{81k} & \textbf{76.62\%} \\
\bottomrule
\end{tabular}
\caption{Comparison of PCA-constrained and standard Transformer architectures on 20 Newsgroups.}
\label{tab:pca_vs_transformer}
\end{table}

\subsection{Case 3: PCA-Guided GPT-style Decoder}

The TinyGPTDecoder is trained to reconstruct token sequences from PCA-compressed MiniLM embeddings with input dimensionality $d = 70$. Despite the reduced input space, the model preserves over 97\% cosine similarity between generated and original embeddings.

The model contains fewer than 3.8 million parameters—less than 17\% of GPT-2’s parameter count—while generating fluent, syntactically plausible text.

\begin{figure}[h]
    \centering
    \includegraphics[width=0.9\linewidth]{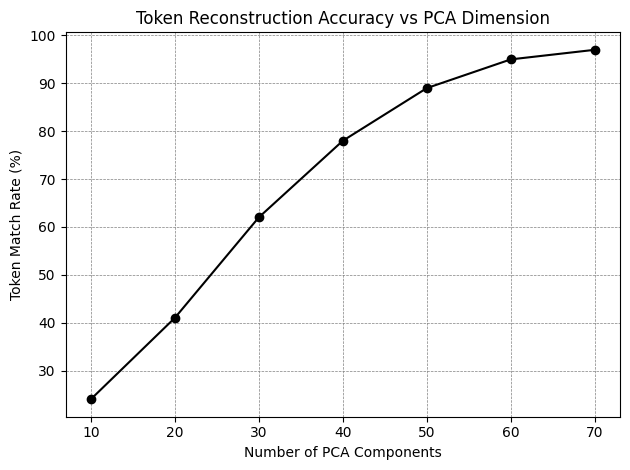}
    \caption{Token-level reconstruction accuracy as a function of PCA dimensionality. The plot shows the percentage of correctly matched tokens after PCA projection and reconstruction using MiniLM embeddings. Each token in the test sentence is mapped to the most similar token in the original embedding set based on cosine similarity. The results demonstrate a rapid increase in reconstruction fidelity as more principal components are retained. Notably, over 90\% of the tokens can be recovered using only 60 components, illustrating the strong compressibility and redundancy within transformer embeddings.}
    \label{fig:accuracy_vs_data}
\end{figure}

To evaluate how well the structure of token embeddings can be preserved under dimensionality reduction, we perform a reconstruction experiment using PCA. Starting from original MiniLM embeddings of 100 tokens, we project both a reference and a test sentence into a common PCA subspace of varying dimensionality. After reconstruction, each token is matched against the original by selecting the closest embedding (in cosine similarity) from the original sequence. The proportion of exact token matches is then measured as a function of the number of PCA components retained, see Figure \ref{fig:accuracy_vs_data}.

\begin{table}[h]
\centering
\caption{TinyGPTDecoder architecture configuration}
\begin{tabular}{l l}
\toprule
\textbf{Component}            & \textbf{Specification} \\
\midrule
Input dimension               & $d = 70$ (PCA-reduced MiniLM) \\
Sequence length               & $T = 100$ tokens \\
Positional encoding           & Sinusoidal (non-trainable) \\
Transformer decoder layers    & 2 \\
Attention heads               & 2 per layer \\
Feedforward hidden size       & 256 \\
Output layer                  & Linear projection to vocabulary ($d \rightarrow 50257$) \\
Vocabulary                    & GPT-2 tokenizer ($|V| = 50257$) \\
\bottomrule
\end{tabular}
\end{table}

To quantify the parameter efficiency gained through dimensionality reduction, we compare the total number of trainable parameters in a Transformer decoder model as a function of network depth. Two configurations are evaluated: one using PCA-compressed embeddings with 70 dimensions, and one using the original full embedding size of 384 dimensions, as found in models such as MiniLM.

Both models share the same decoder architecture and vocabulary size, allowing for a direct comparison. However, due to the quadratic growth of self-attention and feedforward layers with respect to the hidden size, the model with full-dimensional input exhibits significantly higher parameter counts, even for shallow networks.

The following figure \ref{fig:model_size_vs_depth} illustrates this relationship using dual y-axes to better visualize the parameter magnitudes for both configurations.

\begin{figure}[h]
    \centering
    \includegraphics[width=0.9\linewidth]{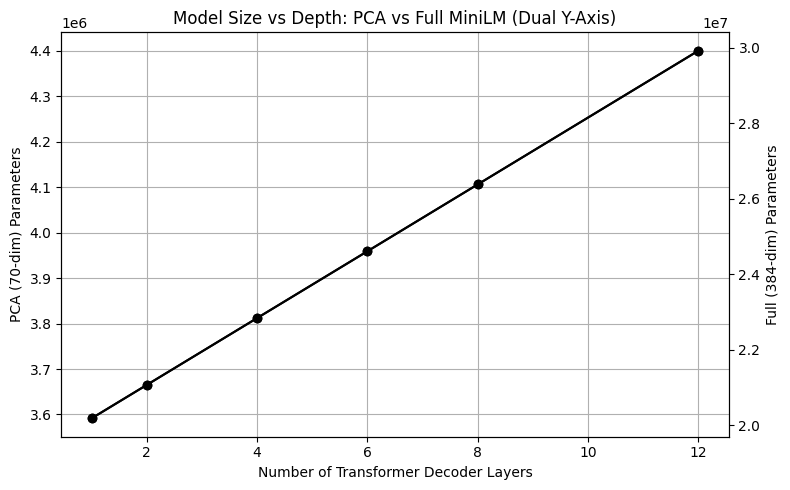}
    \caption{
        Parameter count as a function of Transformer decoder depth, comparing a PCA-compressed model (70-dimensional input) with a full embedding model (384-dimensional input). 
        Both configurations use varying numbers of decoder layers (1 to 12), and the parameter counts include the attention layers, feedforward blocks, and the final output projection.
        The left y-axis corresponds to the compressed model using PCA-reduced input dimensions, while the right y-axis corresponds to the full MiniLM-style model.
        Despite the large difference in input dimensionality, the PCA variant retains a linear scaling with depth and offers a drastic reduction in total parameters---almost an order of magnitude---while still preserving the decoder architecture.
    }
    \label{fig:model_size_vs_depth}
\end{figure}

\subsection{Semantic Compression as Genetic Blueprint}

The proposed PCA-constrained transformer decoding process bears a striking conceptual similarity to biological DNA replication. In our model, the PCA projection acts as a latent ``semantic genome''—a compressed, information-rich representation derived from a base sequence of reference embeddings. This base is constructed from a single, representative information source (e.g., a known input text), analogous to a parent DNA strand encoding the blueprint for future copies.

Just as biological systems replicate complex organisms using a finite alphabet and minimal redundancy, our model reconstructs meaningful output sequences from a drastically reduced input dimensionality (e.g., 70 instead of 384). The transformer decoder plays the role of the cellular replication machinery, expanding the compact PCA basis into full linguistic structures. The result is a ``daughter strand'' of tokens that closely mirrors the original semantic content, despite never having seen the full-dimensional embeddings during training.

This perspective emphasizes not only the computational efficiency of the method—achieving similar output with approximately 1/20th the parameter count—but also its philosophical elegance: semantic regeneration from compressed code, governed by structure rather than surface form.

\begin{figure}[h]
    \centering
    \includegraphics[width=0.9\linewidth]{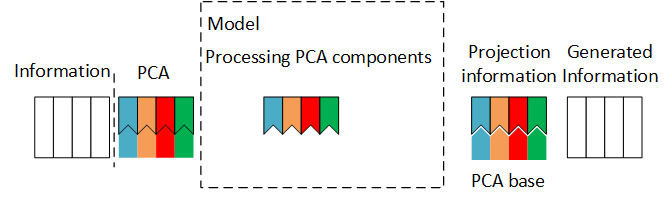}
    \caption{Schematic illustration of PCA-based data transformation and reconstruction, analogous to DNA replication. Raw information is first compressed into PCA components, which are processed by the model. The output is then projected back onto the original PCA basis to reconstruct the generated information. This mirrors how a DNA base strand guides the synthesis of a new complementary strand.}
    \label{fig:pca_dna_analogy}
\end{figure}

An intuitive analogy to DNA replication can be used to conceptualize the PCA-based information processing pipeline. In this view, the original data (information) is compressed into a smaller set of principal components, analogous to a base sequence (template DNA). These components are then processed by a model that operates solely on the compressed representation, similar to how cellular machinery processes nucleotides. Finally, the output is projected back using the same PCA basis, reconstructing the original information in a generative manner. This can be interpreted as synthesizing a daughter strand using the original base as reference, see Figure \ref{fig:pca_dna_analogy}.

\subsection*{Performance Trade-offs and Computational Observations}

While our PCA-constrained architectures achieve substantial compression with competitive accuracy, we observe that the computational overhead of structured PCA must be carefully considered—particularly in the case of small-scale data such as MNIST. In our segment-wise PCA setup, the entire data manifold is projected relative to a fixed reference image. For low-resolution inputs, this transformation introduces a processing cost that often outweighs any benefit in training speed. The bottleneck stems not from the number of training samples $N$, but from the PCA computations themselves, which are currently executed on the CPU and repeated across segments.

This overhead becomes a limiting factor: although the model is compact (e.g., 840 parameters), the preprocessing time scales nearly linearly due to repeated PCA transformations. As such, for small inputs, the gains from compression can be offset by transformation costs—especially when applied at scale.

In contrast, when the same structured PCA method is applied to higher-dimensional data, such as token embeddings in transformer-based language models, the efficiency benefits become more pronounced. The PCA step becomes highly parallelizable, and when executed on a GPU, it leads to considerable speedup in both data preparation and training. These findings suggest that while segment-wise PCA is computationally elegant, its practical value increases with input dimensionality. We provide a comparative runtime analysis of CPU and GPU-based PCA for textual data in Table~\ref{tab:fc3_pca_results}.

\begin{table}[h]
\centering
\begin{tabular}{ll}
\toprule
\textbf{Metric} & \textbf{Value} \\
\midrule
Architecture & 3-layer fully connected \\
Input & 84-dimensional PCA (polar MNIST) \\
Number of parameters & 1,283 \\
Training samples & 10,000 \\
Validation samples & 1,000 \\
Training time & 108.0124 seconds (incl. PCA preprocessing) \\
Test accuracy & 99.2\% \\
\bottomrule
\end{tabular}
\caption{Performance and training characteristics of a 3-layer fully connected model trained on PCA-compressed MNIST data.}
\label{tab:fc3_pca_results}
\end{table}

\section{Conclusion}
This paper demonstrates that Principal Component Analysis, when applied in a structured manner, can serve as a general-purpose compression technique across diverse neural architectures and data modalities. By aligning input representations with the intrinsic topology of the data—whether in images, token embeddings, or sequences—we show that neural models can achieve strong performance with drastically reduced parameter counts.

Each of the three case studies highlights a different facet of this principle:

\begin{itemize}
    \item For image classification, PCA on polar-transformed inputs enabled a one-layer model with only 840 parameters to reach high accuracy.
    \item For text classification, token-wise PCA reduced MiniLM embeddings from 384 to 70 dimensions, supporting an 81k-parameter transformer with competitive performance.
    \item For sequence generation, PCA-compressed embeddings preserved over 97\% semantic similarity in a GPT-style decoder using less than 17\% of the standard parameter count.
\end{itemize}

Together, these results suggest that compression at the data representation level can be more impactful than downstream pruning or architectural simplification. Instead of building ever-larger models, we advocate for front-loading semantic compression using techniques like PCA to match model capacity with information content.

Future work may extend this principle by integrating PCA with knowledge distillation, hardware-aware inference pipelines, or layer-wise component decomposition. More broadly, the approach opens a path toward efficient, interpretable, and structurally aware neural systems that learn more by seeing less.

\section*{Future Work}

Building on the findings of this study, future work will explore the application of structured PCA compression in event-driven computer vision. Specifically, we aim to extend the segment-wise PCA framework to event-based cameras, where asynchronous, sparse data streams challenge conventional frame-based processing pipelines.

Our upcoming work will investigate how temporal PCA applied over event sequences can serve as a preprocessing step to extract meaningful spatiotemporal structure prior to object detection. The goal is to construct lightweight yet responsive models that can operate efficiently on event streams while preserving detection accuracy.

In particular, we will address limitations of conventional architectures such as YOLO and EfficientDet in the context of event-driven tracking and detection. These models, while powerful on frame-based inputs, often struggle with the sparse and asynchronous nature of event data. By restructuring the input representation through sequence-aware PCA, we aim to enable more compact and temporally adaptive object detection models suitable for neuromorphic vision systems.

\section*{Acknowledgements}
The author acknowledges his independence in conducting this research, and the complete absence of project funding, deadlines, or supervision.

\section*{Code Availability}
We see great potential in the use of piecewise PCA to further accelerate and simplify existing machine learning pipelines. As a gesture of openness—and perhaps redemption for all the redundant data we have stripped away—we have prepared a set of Jupyter notebooks covering the three case studies presented in this paper.

These notebooks include implementations for image classification on polar-transformed MNIST, text classification using token-wise PCA compression, and sequence generation with PCA-reduced embeddings. We hope they serve as both a practical resource and an inspiration for continued exploration into structure-first, information-efficient modeling.
\url{https://github.com/mnbe1973/PCA\_LLM}
\bibliographystyle{elsarticle-num}  
\bibliography{references}

\end{document}